\pdfoutput = 1
\documentclass[%
 reprint,
 prd,
 amsmath,amssymb,
 aps,
 superscriptaddress,
 floatfix
]{revtex4-1}

\usepackage{dcolumn}
\usepackage{bm}
\usepackage{ulem}
\usepackage{multirow}
\usepackage{placeins}
\usepackage{algorithmic}
\usepackage{color}
\usepackage{comment}
\usepackage{hyperref}

\usepackage{array}
\newcolumntype{P}[1]{>{\centering\arraybackslash}p{#1}}
\begin{document}

\title{Superradiance of anyons}
\author{Vishnulal Cheriyodathillathu}\thanks{lal23@iisertvm.ac.in}
\affiliation{School of Physics, Indian Institute of Science Education and Research Thiruvananthapuram, Maruthamala PO, Vithura, Thiruvananthapuram 695551, Kerala, India}
\author{Saurya Das}\thanks{saurya.das@uleth.ca}
\affiliation{Theoretical Physics Group and Quantum Alberta, Department of Physics and Astronomy, University of Lethbridge 4401 University Drive, Lethbridge, Alberta T1K 3M4, Canada}
\author{Soumen Basak}\thanks{sbasak@iisertvm.ac.in}
\affiliation{School of Physics, Indian Institute of Science Education and Research Thiruvananthapuram, Maruthamala PO, Vithura, Thiruvananthapuram 695551, Kerala, India}
\date{\today}

\begin{abstract}
In this paper, we investigate superradiance of anyons from a $(2+1)-$ dimensional Bañados, Teitelboim, and Zanelli (BTZ) black hole. Our analysis demonstrates that the superradiance condition for anyons mirrors that of a neutral scalar field within a BTZ black hole. Furthermore, we explore the feasibility of observing this phenomenon in analogue black holes and formulate the corresponding superradiance condition for acoustic black holes.
%
\end{abstract}

\maketitle

\setlength{\columnsep}{0.5cm}

\section{Introduction}
A stationary black hole can be fully characterized by three parameters: its mass \(M\), charge \(Q\), and angular momentum \(J\). The presence of angular momentum leads to a fascinating phenomenon known as "superradiance," where rotational energy is extracted and radiated to infinity, resulting in a reduction of both the mass and angular momentum of a spinning black hole \cite{1971ZhPmR..14..270Z}. This process is the field analogue of the Penrose process \cite{Penrose:1971uk}, in which energy extraction occurs through mechanical scattering. Superradiance can also occur in charged black holes, where charge is extracted via interactions with quantum fields. Black hole superradiance is extensively studied across various spacetimes and particle types \cite{Martellini:1977qf,PhysRevD.83.044026, ZOUROS1979139, PhysRevD.86.104017, PhysRevLett.109.131102, PhysRevD.96.035019, PhysRevD.88.023514,PhysRevD.81.123530,PhysRevD.91.124026}. However, the occurrence of superradiance in the context of anyons for rotating black holes remains an open question. This work aims to address this gap by exploring superradiance specifically in the context of $(2+1)-$ dimensional rotating black holes.

Anyons are particles hypothesized to exist exclusively in \((2+1)\)-dimensional spacetime, which are neither bosons nor fermions but instead obey fractional statistics \cite{Rao1992AnAP, Sen:1993qc}. These particles are not merely theoretical constructs, as there is experimental evidence supporting their existence \cite{Bartolomei:2020qfr}. The practical applications of anyons, particularly in quantum computation\cite{KITAEV20032}, underscore the importance of studying these particles.

This research is driven by two primary motivations. First, we aim to test the existence of anyons and thereby gain a deeper understanding of their properties. Confirmation of our results within the proposed experimental setup would indicate the presence of anyons.
Consequently, studying superradiance for anyons from a \((2+1)\)-dimensional analogue
black hole offers a new avenue for exploring these particles. Second, we seek to examine the properties of quantum fields in the presence of a black hole horizon, resulting in superradiance. A deeper understanding of these quantum phenomena will ultimately contribute to our knowledge of the quantum nature of black holes.

The primary objective of this paper is to investigate the superradiance effect of anyons from BTZ and analogue black holes and to interpret the corresponding results. In the following section, we provide a brief overview of anyons, which obey fractional statistics, with their spin taking any value between zero and one in units of \(\hbar\). Section \ref{QNMscalar} reviews the superradiance phenomenon for a massive scalar field from a BTZ black hole. In Section \ref{supranyn}, we examine the superradiance of anyons from a BTZ black hole. Section \ref{experimentalset-up} explores potential systems where this phenomenon could be observed in a laboratory setting and discusses experimental results that may have already provided verification. Finally, we summarize our findings and conclude in Section \ref{conclusionsection}.
\section{Anyons}
\label{anyons}
It is generally accepted that particles in three-dimensional space, or \((3+1)\)-dimensional spacetime, are classified as either bosons or fermions. These particles are distinguished by their integral or half-integral spins (in units of \(\hbar\)), adhering to Bose-Einstein and Fermi-Dirac statistics, respectively, and are described by symmetric or anti-symmetric wavefunctions under the exchange of particles. However, the situation becomes significantly less restrictive in \((2+1)\)-dimensional spacetime, where a continuous range of statistics is permitted \cite{Sen:1993qc}. This can be demonstrated straightforwardly by examining the wavefunction's behavior for two identical particles in \((2+1)\)-dimensional spacetime.

Let \(\psi(r)\) be the wave function of a system comprising the two particles, subject to the condition that \(\psi(r) \neq 0\) for \(r > a\) (the so-called ``hard-core condition"), where \(\vec{r}_1\) and \(\vec{r}_2\) are the position vectors of the two particles, and \(\vec{r} \equiv \vec{r}_1 - \vec{r}_2\) is the relative position vector. The configuration space of these particles is therefore the two-dimensional \((x,y)\)-plane with a disc of radius \(a\) removed.

By defining a complex coordinate \(z = x + i y\) and applying the transformation \(z \rightarrow z\,e^{2\pi i}\), which returns the particle to its original position, the wave function should remain invariant, except for a phase factor. This can be expressed as:
\begin{eqnarray}
   \psi(z e^{i2\pi}, z^{*} e^{-i2\pi}) = e^{i 2 \pi \alpha} \psi(z, z^{*}) ~,
   \label{Eq1}
\end{eqnarray}
where \(\alpha\) is a real parameter. Similarly, when interchanging the two particles by transforming \(z \rightarrow z e^{i \pi}\), we obtain:
\begin{eqnarray}
   \psi(z e^{i\pi}, z^{*} e^{-i\pi}) = e^{i \pi \alpha} \psi(z, z^{*})~,
   \label{Eq2}
\end{eqnarray}
where \(\alpha\) takes the value of \(0\) for bosons and \(1\) for fermions. However, in \((2+1)\)-dimensions, \(\alpha\) can be any real number between \(0\) and \(1\).

To see this, consider a system of two identical particles in \((3+1)\)-dimensions. To return to the original configuration, two consecutive interchanges of the particles' locations are required. All such trajectories are topologically equivalent. However, in \((2+1)\)-dimensions, after one interchange, an additional winding of one particle around the other is needed to return to the initial configuration. Unlike in \((3+1)\)-dimensions, these two trajectories are distinct and cannot be deformed into each other. Therefore, they can be associated with two distinct topological phases, which can take any value between \(0\) and \(1\). This demonstrates that in two spatial dimensions, a particle may possess statistics different from the standard Bose-Einstein and Fermi-Dirac statistics, termed as fractional statistics. Particles following fractional statistics are called anyons. The primary focus of this work is to describe the superradiance of these particles.

\section{Superradiance of massive scalar field from BTZ black hole}
\label{QNMscalar}
In this section, we investigate a massive scalar field \(\Phi\) in \((2+1)\)-dimensional spacetime, governed by the Klein-Gordon equation:
\begin{equation}
   \left( \frac{1}{\sqrt{-g}}\partial_{\mu}\left(\sqrt{-g}g^{\mu \nu}\partial_{\nu}\right)\,-\,\epsilon R_{3}\,-\,\mu^{2} \right)\Phi=0~,
    \label{eqmotionscalar}
\end{equation}
where \(R_{3}\) is the three-dimensional Ricci scalar, and \(\epsilon\) denotes its coupling to the scalar field. The curvature is given by \(R_{3}=-\frac{1}{l^{2}}=6 \Lambda\), where \(\Lambda\) is the negative cosmological constant, and \(\mu\) represents the mass of the scalar field. We study this field in the background of a BTZ black hole. The metric for a BTZ black hole, arising from \((2+1)\)-dimensional topological gravity with a negative cosmological constant, is given by \cite{PhysRevLett.69.1849}:
\begin{equation}
ds^{2}=-f(r)dt^{2}+\frac{dr^{2}}{f(r)}+r^{2}(d\phi + N^{\phi}\,dt)^{2}~,
    \label{3.2}
\end{equation}
where
\begin{equation}
    f(r)\,=\,\frac{(r^{2}-r_{+}^{2})(r^{2}-r_{-}^{2})}{l^{2}r^{2}}, \hspace{30pt} N^{\phi}\,=\,\frac{r_{+}r_{-}}{l\,r^{2}}.
\end{equation}
The radii \(r_{+}\) and \(r_{-}\) correspond to the outer and inner horizons, with their relation to the black hole parameters expressed as:
\begin{equation}
    r_{\pm}\,=\,\sqrt{2G_{N}(M+J)}\pm\sqrt{2G_{N}(M-J)} ~,
\end{equation}
Here, \(M\) and \(J\) represent the mass and angular momentum of the rotating BTZ black hole, respectively and \(G_N\) is Newton's constant in \((2+1)\)-dimensions. 

To ensure the presence of superradiance and the existence of two event horizons, the condition \(\mid J \mid < M\) must hold. Black holes satisfying this criterion are referred to as nonextremal black holes. The angular velocity at the event horizon is given by \(\Omega_{H}\,=\,-N^{\phi}(r_{+})\).

Given the symmetry of the metric, we can assume the following ansatz for the massive scalar field:
\begin{equation}
    \Phi(t,r,\phi)\,=\,\frac{R(r)}{\sqrt{r}}\,e^{-i\omega t+im\phi},
    \label{3.5}
\end{equation}
where \(\omega\) is the frequency, and \(m\) is the angular momentum quantum number. Substituting this ansatz into Eq.~(\ref{eqmotionscalar}) results in a differential equation for the radial function:
\begin{equation}
    \frac{d^{2}R(r)}{dr^{* 2}} + V_{eff}\,R(r)\,=\,0~,
    \label{diffR1}
\end{equation}
where the effective potential \(V_{eff}(r)\) is given by:
\begin{equation}
    V_{eff}(r) = (\omega + mN^{\phi})^{2} - f(r)\left(\frac{m^{2}}{r^{2}} + \mu^{'2} + \frac{f^{'}}{2r} - \frac{3f}{4r^{2}}\right)~,
    \label{pot1}
\end{equation}
where $\mu^{'2}= \epsilon R_{3} + \mu^{2}$. The above equation can be solved at the event horizon and at infinity. At the event horizon, where the metric function vanishes (\(f(r_{+})=0\)), Eq.~(\ref{diffR1}) simplifies to:
\begin{equation}
    {\frac{d^{2}R}{dr^{* 2}}}|_{r=r_+}\, + \,\omega_{H}^{2}\,R(r_{+})\,=\,0~,
    \label{diffR2.1}
\end{equation}
where
\begin{equation}
    \omega_{H}\,=\,\omega\,-\,m\Omega_{H},
\end{equation}
and \(\Omega_{H}\) is the angular velocity at the event horizon. The behavior of the radial function near the event horizon, following from Eq.~(\ref{diffR2.1}), is:
\begin{equation}
    R(r\approx r_{+}) \sim A_{\omega m}e^{ i \omega_{H} r^{*}} + B_{\omega m}e^{- i \omega_{H} r^{*}}.
    \label{nh1}
\end{equation}

A similar procedure can be employed to determine the behavior of the radial function at infinity. At large radial distances, the differential equation (\ref{diffR1}) takes the form:
\begin{equation}
    {\frac{d^{2}R(r)}{dr^{*2}}} - \left(\frac{r^{2}\mu^{'2}}{l^{2}} + \frac{r^{2}}{4l^{4}}\right)R(r)\,=\,0~,
\label{de10}
\end{equation}
where $r_{*}$ is the tortoise coordinate defined by:
\begin{equation}
    dr_{*}\,=\,\frac{1}{f(r)}dr \hspace{10pt}  \Rightarrow \hspace{10pt} r_{*}\,=\,-\frac{l^{2}}{r}~.
    \label{tortoise}
\end{equation}
With this transformation, Eq.~(\ref{de10}) simplifies to:
\begin{equation}
    \frac{d^{2}R(r_{*})}{dr^{* 2}} - \frac{k^{2}}{r_{*}^{2}}\,R(r_{*})\,=\,0~,
    \label{diffR2}
\end{equation}
where
\begin{equation}
    k^{2}\,=\,\mu^{'2}l^{2}+\frac{1}{4}~.
\end{equation}
The solution to this equation determines the behavior of the radial function at infinity:
\begin{equation}
    R_{\omega m}\,\sim\,r_{*}^{p}\,\sim\,r^{-p}~,
\end{equation}
where we have used Eq.~(\ref{tortoise}), and \(p\) is given by:
\begin{equation}
    p\,=\,\frac{1}{2}\left(1\pm\sqrt{1+4k^{2}}\right)~.
\end{equation}

It can be observed that for \(4k^{2}>-1\), \(p\) is real, leading to the following solution:
\begin{equation}
    R_{\omega m}(r)\,=\,C_{\omega m}r^{-\frac{1}{2}\left(1+\sqrt{1+4k^{2}}\right)} + D_{\omega m}r^{-\frac{1}{2}\left(1-\sqrt{1+4k^{2}}\right)}~.
    \label{sol}
\end{equation}
The second term in Eq.~(\ref{sol}) represents a radial function that is not square integrable at infinity when \(4k^{2} > 0\), leading to the condition \(D_{\omega m}=0\). In this scenario, no boundary conditions can be imposed on the scalar field at infinity. For \( -1 < 4k^{2} < 0\), both terms in Eq.~(\ref{sol}) are square integrable, allowing the choice of Dirichlet boundary conditions (\(D_{\omega m}=0\)) or Neumann boundary conditions (\(C_{\omega m}=0\)). If both coefficients \(D_{\omega m}\) and \(C_{\omega m}\) are nonzero, Robin boundary conditions can be imposed. 

We now check the existence of superradiant modes under these different boundary conditions. Here we find the energy flux across the event horizon, using the Eddington-Finkelstein (EF) coordinates which is regular across the horizon, defined as
\begin{equation}
    dv=dt+ \frac{1}{f(r)}dr   \hspace{15pt}  d\phi_{n}=d\phi-\frac{N^{\phi}(r)}{f(r)}dr~.
    \label{EFcord}
\end{equation}
By considering the solution (Eq.(\ref{nh1})), we choose the following ansatz,
\begin{equation}
    \Phi_{\omega m} = \frac{B_{\omega m}}{\sqrt{r_{H}}}e^{-i(\omega \nu - m \phi_{n})}~.
\end{equation} 
The energy flux across the horizon as follows\cite{DAPPIAGGI2018146, PhysRevD.67.084013},
\begin{equation}
    \mathcal{F}_{E}=\int_{0}^{2\pi}d\phi_{n}r_{H}\chi_{\mu}T^{\mu}_{\nu}\xi^{\nu}
    \label{scalarint}
\end{equation}
where $\chi$ and $\xi$ are the generators of the horizon,
\begin{equation}
    \chi = \partial_{\nu} + \Omega_{H} \partial_{\phi_{n}}  \hspace{15pt} \chi_{\nu}=\partial_{\nu}~,
    \label{generators}
\end{equation}
and $T_{\mu \nu}$ is the stress energy tensor of the scalar field. Evaluating the integral in Eq.(\ref{scalarint}), gives the following result,
\begin{equation}
     \mathcal{F}_{E}= F \, \left(\mathcal{R}(\omega)\left[\mathcal{R}(\omega)-\frac{m\Omega}{l}\right]+ \mathcal{I}(\omega)^{2}\right)
     \label{neutral}
\end{equation}
where,
\begin{equation}
    F= 2\pi r_{H}\mid B_{\omega m} \mid^{2}e^{2\nu \mathcal{I}(\omega)}
\end{equation}
It is easy to see that energy flux across the horizon is negative when $\mathcal{R}(\omega)\left[\mathcal{R}(\omega)-m\Omega_{H}\right]+ \mathcal{I}(\omega)^{2} < 0$, signifying outgoing energy from black holes towards its exterior region, which are the superradiant modes.
\section{Superradiance of anyons from BTZ black hole}
\label{supranyn}
In this section, we will extend the approach of the previous section to an anyonic field $\Phi$. Consider the Lagrangian \cite{Rao1992AnAP},
\begin{multline}
    L=-\frac{1}{4}F_{\mu\nu}F^{\mu\nu}+\frac{1}{2}(\partial_{\mu}-iqA_{\mu})\Phi^{*}\left(\partial^{\mu}+iqA^{\mu}\right)\Phi\\
    -c_{4}\left(\Phi \Phi^{*}-\frac{c_{2}}{2c_{4}}\right)^{2}+\frac{\mu}{4}\epsilon_{\mu\nu\alpha}F^{\mu\nu}A^{\alpha}~.
    \label{4.1}
\end{multline}
The above represents an Abelian Higgs model with a Chern-Simons
(CS) term, whose classical solutions give rise to anyonic excitations. 
Here, $F_{\mu\nu}$ is the electromagnetic field tensor, $A_{\mu}$ is the gauge field and, $c_{2}$  and $c_{4}$ real constants. The parameter $\mu$ has the dimensions of a mass and is the gauge invariant mass term for the
gauge field. This gives the equation of motion of anyonic field as follows,
\begin{multline}
g^{\mu\nu}\nabla_{\mu}\nabla_{\nu}\Phi+2iqg^{\mu \nu}A_{\mu}\partial_{\nu}\Phi\\ 
-\left(2c_2+q^2g^{\mu \nu} A_{\mu} A_{\nu}\right) \Phi + 4 c_4 \Phi^* \Phi^2 = 0~.
\label{4.2}
\end{multline}
Here we consider the charged generalization of the BTZ metric\cite{PhysRevD.61.104013}:\\
\begin{equation}
    ds^{2}=-f(r)\frac{r^{2}}{R(r)^{2}}dt^{2}+\frac{1}{f(r)}dr^{2}+R(r)^{2}[d\phi+N^{\phi}(r)dt]^{2}
\end{equation}
where,
\begin{equation}
    f(r)=\frac{r^{2}}{l^{2}}-M-\frac{Q^{2}}{2}\ln\left({\frac{r}{r_{H}}}\right)
\end{equation}
\begin{equation}
    R(r)^{2}=r^{2}+\frac{\Omega^{2} l^{2}}{1-\Omega^{2}}\left[M+ \frac{Q^{2}}{2} \ln\left({\frac{r}{r_{H}}}\right)  \right]~,
\end{equation}
\begin{equation}
    N^{\phi}(r)=-\frac{\Omega l}{(1-\Omega^{2})R(r)^{2}}\left[M + \frac{Q^2}{2} \ln\left({\frac{r}{r_{H}}}\right)   \right]~,
\end{equation}
where $M$, $Q$ and $\Omega$ $\in (0,1]$ are constants. The mass $\tilde{M}$, angular momentum $\tilde{J}$ and charge $\tilde{Q}$ of the black hole are given in terms of these constant parameters as,
\begin{equation}
    \tilde{M}=\frac{1}{1-\Omega^{2}}\left[M(1+\Omega^{2})-\frac{1}{2}Q^{2}\Omega^{2}\right]
\end{equation}
\begin{equation}
    \tilde{J}=\frac{2\Omega}{1-\Omega^{2}}\left[M-\frac{1}{4}Q^{2}\right]    \hspace{20pt} \tilde{Q}=\frac{Q}{\sqrt{1-\Omega^{2}}}
\end{equation}
Similar to the case of scalar fields in the previous section, we are interested to calculate the energy flux across the event horizon of the black hole. In order to achieve this, we consider a coordinate transformation to make the metric regular at the event horizon as we did earlier in Eq.(\ref{EFcord}),
\begin{equation}
    \nu=t+\frac{r_{*} }{\sqrt{1-\Omega^{2}}}  \hspace{25pt}  \phi_{n}\,=\,\phi-\frac{\Omega r_{*}}{l\sqrt{1-\Omega^{2}}}~,
\end{equation}
where $r_{*}$ is the tortoise coordinate defined as 
\begin{equation}
    \frac{dr_{*}}{dr}=\frac{1}{f(r)}~.
\end{equation}
In addition to this, we use the following ansatz for the anyonic field to obtain a differential equation for the radial part in tortoise coordinate.
\begin{equation}
    \Phi_{\omega m}= \frac{1}{\sqrt{r}}X(r)e^{-i\omega t+ i m \phi} ~,
\end{equation}
In this ansatz, $\omega$ is the frequency of the anyons which is in general complex function, and $m$ is the azimuthal quantum number. 

Inserting this ansatz into Eq.(\ref{4.2}) and setting the constants $c_{2}=c_{4}=0$ and the gauge field $A_{\mu}=-Q\ln({\frac{r}{r_{0}}}) \delta^{t}_{\mu}$, which is just the Coulomb's law in $(2+1)-$dimensions, we get the desired differential equation.
\begin{equation}
    \left[ \frac{d^{2}}{dr_{*}^{2}} + V_{\omega m}(r) \right] X_{\omega m}(r) =0~,
    \label{diffeq4}
\end{equation}
where $V_{\omega m}(r)$ is the effective potential. 
\begin{multline}
    V_{\omega m}(r)=\left[ \frac{\omega - m\Omega l^{-1}}{\sqrt{1-\Omega^{2}}}-qQ\ln{\left(\frac{r}{r_{H}}\right)}   \right]^{2}\\ + \frac{f(r)^{2}}{4r^{2}}-\frac{f^{'}(r) f(r)}{2r}-\frac{(\omega \Omega - ml^{-1})^{2}l^{2}f(r)}{(1-\Omega^{2})r^{2}}~.
\end{multline}
Near the event horizon, for which $r \rightarrow r_{H}$ ($r_{*} \rightarrow -\infty$) the above differential equation reduces to the following form,
\begin{equation}
    {\frac{d^{2}X_{\omega m}}{dr^{* 2}}}|_{r=r_H}\,
    +\,\omega_{H}^{2}\,X_{\omega m}(r_{H})\,=\,0~,
    \label{diffR2.11}
\end{equation}
where,
\begin{equation}
    \omega_{H}\,=\,\frac{\omega\,+\,m\Omega_{H}}{\sqrt{1-\Omega^{2}}}~,
\end{equation}
and $\Omega_{H}=-\Omega l^{-1}$, is the angular velocity at the event horizon. 

The most general solution to this differential equation is given by,
\begin{equation}
    X_{\omega m}(r) \sim A_{\omega m}e^{i \omega_{H} r_{*}} + B_{\omega m} e^{-i \omega_{H} r_{*}}
    \label{solinf}
\end{equation}
$A_{\omega m}$ and $B_{\omega m}$ are the complex constants. 
Following the same procedure we find the behaviour of radial function at $r \rightarrow \infty$. In this limit, Eq.(\ref{diffeq4}) can be reduced to the following form,
\begin{equation}
    \left[\frac{d^{2}}{dr_{*}^{2}} - \frac{3}{4r_{*}^{2}}\right]X_{\omega m}(r)=0~.
\end{equation}
The most general solution to this differential equation is given by,
\begin{equation}
    X_{\omega m} \sim C_{\omega m} r^{-\frac{3}{2}} + D_{\omega m}  r^{\frac{1}{2}}
\end{equation}
$C_{\omega m}$ and $D_{\omega m}$ are the complex constants. 
The second term in the above solution diverges at infinity. In order to exclude this unphysical solution from our analysis, we set $D_{\omega m}=0$.

To find the condition for superradiance of anyons, we compute the energy flux across the event horizon using the near horizon solution. We need suitable coordinates which are regular across the horizon for which we use the EF coordinates. The ingoing null coordinate $v$ and the angular coordinate $\phi$ are defined as follows,
\begin{equation}
   dv=dt+\frac{1}{r} \frac{R(r)}{f(r)}dr   \hspace{15pt}  d\phi_{n}=d\phi-\frac{R(r)}{r}\frac{N^{\phi}(r)}{f(r)}dr~.
\end{equation}

Using these coordinates, we can write the behaviour of radial function for an ingoing mode at the event horizon as follows (Eq.(\ref{solinf})),
\begin{equation}
    \Phi_{\omega m} = \frac{B_{\omega m}}{\sqrt{r_{H}}}e^{-i(\omega \nu - m \phi_{n})}
\end{equation}
The energy flux across the event horizon of the black hole for this solution is given by,
\begin{equation}
    \mathcal{F}_{E}=\int_{0}^{2\pi}d\phi_{n}r_{H}\chi_{\mu}T^{\mu}_{\nu}\xi^{\nu}
\end{equation}
where $T^{\mu}_{\nu}$ is the energy momentum tensor for the anyonic field and $\chi_{\mu}$, $\xi^{\nu}$ are the killing vectors as in Eq.(\ref{generators}).
\begin{multline}
    T_{\beta \nu} =  F^{\mu}_{\beta} F_{\mu \nu}-(D_{\beta}\Phi)^{*}D_{\nu}\Phi\\-\epsilon_{\beta \delta \chi}(2F_{\nu \mu}A_{\alpha}+F_{\mu \alpha}A_{\nu})g^{\chi \alpha}g^{\delta \mu} \\- \frac{1}{2}g_{\beta \nu} \left(\frac{1}{2}F_{\alpha \mu} F^{\alpha \mu} - (D_{\mu}\Phi)^{*}D^{\mu}\Phi - \frac{\mu}{2} \epsilon^{\mu \delta \alpha} F_{\mu \delta} A_{\alpha} \right)
\end{multline}
Considering the relevant component of energy momentum tensor and calculating the above integral we find the expression of the flux across the horizon,
\begin{equation}
     \mathcal{F}_{E}= F \, \left(\mathcal{R}(\omega)\left[\mathcal{R}(\omega)-\frac{m\Omega}{l}\right]+ \mathcal{I}(\omega)^{2}\right)
\end{equation}
where,
\begin{equation}
    F= 2\pi r_{H}\mid B_{\omega m} \mid^{2}e^{2\nu \mathcal{I}(\omega)} ~.
\end{equation}
The energy flux across the horizon needs to be negative for superradiance to happen and this occurs when the following condition is satisfied:
\begin{equation}
    \mathcal{R}(\omega)\left[\mathcal{R}(\omega)-m\Omega_{H}\right]+ \mathcal{I}(\omega)^{2} < 0 ~.
\end{equation}
The above result is identical 
 to the condition for the superradiance of neutral scalar field from the BTZ black hole (Eq.(\ref{neutral})). This result also matches with the condition for superradiance for a charged scalar field \cite{Konewko2023ChargeSO}.
\section{Experimental setup- Superradiance of anyons from acoustic black holes}
\label{experimentalset-up}
In this section, we explore the possibility of validating our findings through analog models of gravity. To be specific, we show that anyons display superresonance, which is the counterpart of superradiance in a class of analogue or acoustic black holes, which is basically the amplification of acoustic disturbances due to the scattering from an ergo region of a rotating acoustic black hole. 

In the case of a two-dimensional photon-superfluid system, it has already been shown that the kinematics is governed by the equation of a massless scalar field in the background of an acoustic metric \cite{PhysRevA.78.063804}. We have already discussed in our earlier works, that the equation of motion of this superfluid system, augmented by a set of corrections which may be realizable in the laboratory, governs the dynamics of anyons in the background of the acoustic metric \cite{PhysRevD.104.104011,Cheriyodathillathu:2022fwe}. 
This is our motivation to study superradiance (or, superresonance) from an acoustic black hole in an experimental set-up. 

We choose a system of $(2+1)$-dimensional fluid flow with a sink at the origin. This is known as a 'draining bathtub’ type of fluid flow. In polar coordinates, the velocity potential of the fluid flow takes the following form \cite{Visser:1997ux,SoumenBasak_2003}
\begin{equation}
    \psi(r,\phi)\, =\, A \log(r)\,+\,B\phi~,
\end{equation}
where A and B are two real constants. The corresponding velocity profile is given by,
\begin{equation}
   \vec v\,= \vec\nabla\psi = 
   \,\frac{A}{r}\hat{r}+\frac{B}{r}\hat{\phi}~.
\end{equation}
The corresponding metric of an acoustic black hole can be written as follows,
\begin{multline}
    ds^{2}\,=\,-\left(c^{2}-\frac{A^{2}+B^{2}}{r^{2}} \,\right)dt^{2}\,-\,\frac{2A}{r}\,dr \,dt \\
     \,-\,2B\, d\phi \,dt\,+\,dr^{2}\,+\,r^{2}d\phi^{2}~,
     \label{acoustic1}
\end{multline}
where $c$ is the speed of sound. 

We will first simplify the above line element by switching to another coordinate system, 
\begin{equation}
    dt\,\,\,\,\rightarrow\,\,\,\,dt\,+\,\frac{\mid A \mid r}{r^{2}c^{2}-A^{2}}\,dr
\end{equation}
\begin{equation}
    d\phi\,\,\,\,\rightarrow\,\,\,\,d\phi\,+\,\frac{\mid A \mid B}{r^{2}c^{2}-A^{2}}\,dr ~,
\end{equation}
where the metric component $g_{rt}$ vanishes.

\begin{multline}
    ds^{2}\,=\,-\left(1-\frac{A^{2}+B^{2}}{c^{2}r^{2}} \,\right)dt^{2}\,+\,\left(1-\frac{A^{2}}{c^{2}r^{2}} \,\right)^{-1}dr^{2}\\-\,2\frac{B}{c}\, d\phi \,dt\,+\,r^{2}d\phi^{2}
    \label{metric}
\end{multline}
As in the case of the Kerr black hole, the radius of the ergo region is the solution of $g_{tt}=0$ which gives, $r_{e}=\frac{(A^{2}+B^{2})^{\frac{1}{2}}}{c^{2}}$. Similarly, the solution of the equation $g_{rr}=0$ gives the radius of the event horizon, $r_{h}=\frac{\mid A \mid}{c}$.

Following the procedure summerized in \cite{PhysRevD.104.104011}, we consider anyons in the background of the above metric. For simplicity of calculations, we set the parametes $c_2$ and $c_4$ to $0$. In addition to this, we choose the components of the vector potential as, $A_{\mu}=(0,0, a)$. For these choices, the equation motion of anyons can be written in the following form,
\begin{equation}
g^{\mu\nu}\nabla_{\mu}\nabla_{\nu}\Phi+2iqg^{\mu \nu}A_{\mu}\partial_{\nu}\Phi
-q^2g^{\mu \nu} A_{\mu} A_{\nu} \Phi = 0~.
\label{eqmo}
\end{equation}

We use the following ansatz to solve the above differential equation,
\begin{eqnarray}
    \Phi(r\,,t\,,\theta)\,=\,R(r)\,e^{i\,\omega\,t}\,e^{-i\,m\,\phi}~.
\end{eqnarray}
Inserting the ansatz and metric components into Eq.~(\ref{eqmo}) yields the following differential equation for \( R(r) \):
\begin{multline}
    \frac{1}{r}\left( 1-\frac{A^{2}}{c^{2}r^{2}} \right) \frac{d}{dr}\left(r\left(1-\frac{A^{2}}{c^{2}r^{2}}\right)\frac{d}{dr}\right)R(r) \\
    +\left[ \omega^{2}-\frac{2\omega B(m+aq) }{cr^{2}} \right. \\
    \left.-\frac{(m+aq)^{2}}{r^{2}}\left(1-\frac{A^{2}+B^{2}}{c^{2}r^{2}}\right)\right]R(r)=0~.
\end{multline}

We introduce a coordinate transformation from $r$ to tortoise coordinates, $r_{*}$,
\begin{equation}
    \frac{d}{dr^{*}}=\left( 1-\frac{A^{2}}{r^{2}c^{2}}\right)\frac{d}{dr}~,
\end{equation}
which results in a simpler differential equation for the modified radial part of the ansatz,
\begin{equation}
    \frac{d^{2}G(r_{*})}{dr^{2}_{*}}+\left[Q(r)+\frac{1}{4r^{2}}\left(\frac{dr}{dr_{*}}\right)^{2}- \frac{A^{2}}{r^{4}c^{2}}\left(\frac{dr}{dr_{*}}\right)  \right]G(r_{*})=0~,
    \label{eqnG3}
\end{equation}
where,
\begin{equation}
G(r_*)=\sqrt{r}R(r)
\end{equation}
and 
\begin{equation}
    Q(r)=\omega^{2}-\frac{2B\omega (m+aq) }{cr^{2}} -\frac{(m+aq)^{2}}{r^{2}}\left(1-\frac{A^{2}+B^{2}}{c^{2}r^{2}}\right)~.
\end{equation}
Now we solve the Eq.(\ref{eqnG3}) in the asymptotic regions ($r^{*}\rightarrow +\infty$ and $r^{*}\rightarrow -\infty$ ). Near the horizon, the differential equation (\ref{eqnG3}) reduces to a very simple form,
\begin{equation}
    \frac{d^{2}G(r^{*})}{dr^{*2}}\,+\, (\omega-(m+aq)\Omega_{H})^{2}G(r^{*})=0
\end{equation}
This gives the behaviour of $G(r_*)$ near horizon as follows,
\begin{equation}
    G(r^{*})\,=\,\mathcal{T}\,e^{i(\omega-(m+aq)\Omega_{H})r^{*}}~.
\end{equation}
Similarly, at infinity, Eq.(\ref{eqnG3}) reduces to
\begin{equation}
    \frac{d^{2}G(r^{*})}{dr^{*2}}\,+\, \omega^{2}G(r^{*})=0~,
\end{equation}
This gives the behaviour of $G(r_*)$ at $r\rightarrow +\infty$  as follows,
\begin{equation}
    G(r^{*})\,=\,e^{i\omega r^{*}}+ \mathcal{R} e^{-i \omega r^{*}}~.
\end{equation}
Calculating the Wronskian of the solutions in either case gives the relation between transmission and reflection coefficients,
\begin{equation}
    \mid R \mid^{2}\,=\,1\,-\,\left(\frac{\omega-(m+aq)\Omega_{H}}{\omega}\right)\,\mid T \mid^{2}~.
\end{equation}
From the above expression, it is evident that the reflection coefficient $\mathcal{R}$ is greater than unity when $0<\omega<(m+aq)\Omega_{H}$, which is equivalent to superradiance from a rotating black hole.\\

\section{Conclusion} \label{conclusionsection}

In this paper, we extended the study of superradiance to a novel class of particles, known as anyons. Our analysis led to the conditions for the existence of superradiance from both $(2+1)-$ dimensional BTZ black holes and acoustic black hole. This opens up new avenues for investigating the phenomenon in lower-dimensional systems.

Moreover, our theoretical predictions suggest that these effects may be experimentally validated in laboratory settings. Such an experimental confirmation would not only provide critical insights into the nature of superradiance but also serve as compelling evidence for the existence of anyons — an idea that remains largely theoretical.

In summary, our contributions include the following key points: We provided a comprehensive review of superradiance in neutral scalar fields from BTZ black holes, deriving the necessary conditions for the phenomenon to occur. We then extended the analysis to anyons and found that the conditions for superradiance in the case of anyons from BTZ black holes are identical to those for neutral scalar fields. We further explored the experimental feasibility of observing anyonic superresonance—an analogue to superradiance, and derived the conditions to see this in laboratory-created acoustic black holes, noting that such observations could serve as concrete tests of our theoretical predictions.

Looking ahead, we are optimistic that the effects described here will soon be measurable, providing a deeper understanding of superradiance and contributing to the broader field of quantum gravity and black hole physics.
\vspace{30pt}

\section{Acknowledgement}
This work was supported by the Natural Sciences and Engineering Research Council of Canada. 
Research fellowships from the Council of Scientific and Industrial Research, India supported V.C.\\

\nocite{PhysRevLett.69.1849}\nocite{Sen:1993qc}\nocite{1971ZhPmR..14..270Z}\nocite{PhysRevA.78.063804}\nocite{PhysRevD.104.104011}\nocite{Visser:1997ux}\nocite{Martellini:1977qf}\nocite{KITAEV20032}\nocite{Cheriyodathillathu:2022fwe}
\nocite{Penrose:1971uk}\nocite{PhysRevD.81.123530}\nocite{PhysRevD.83.044026}\nocite{ZOUROS1979139}\nocite{PhysRevD.86.104017}\nocite{PhysRevLett.109.131102}\nocite{PhysRevD.88.023514}\nocite{Rao1992AnAP}\nocite{Bartolomei:2020qfr}\nocite{Brennen2007WhySA}\nocite{DAPPIAGGI2018146}\nocite{PhysRevD.61.104013}\nocite{Konewko2023ChargeSO}\nocite{SoumenBasak_2003}\nocite{PhysRevD.91.124026}\nocite{PhysRevD.67.084013}
\bibliographystyle{ieeetr}
\bibliography{main}

\begin{thebibliography}{10}

\bibitem{1971ZhPmR..14..270Z}
Y.~B. {Zel'Dovich}, ``{Generation of Waves by a Rotating Body},'' {\em ZhETF Pisma Redaktsiiu}, vol.~14, p.~270, Aug. 1971.

\bibitem{Penrose:1971uk}
R.~Penrose and R.~M. Floyd, ``{Extraction of rotational energy from a black hole},'' {\em Nature}, vol.~229, pp.~177--179, 1971.

\bibitem{Martellini:1977qf}
M.~Martellini and A.~Treves, ``{Absence of Superradiance of a Dirac Field in a Kerr Background},'' {\em Phys. Rev. D}, vol.~15, pp.~3060--3061, 1977.

\bibitem{PhysRevD.83.044026}
A.~Arvanitaki and S.~Dubovsky, ``Exploring the string axiverse with precision black hole physics,'' {\em Phys. Rev. D}, vol.~83, p.~044026, Feb 2011.

\bibitem{ZOUROS1979139}
T.~J. Zouros and D.~M. Eardley, ``Instabilities of massive scalar perturbations of a rotating black hole,'' {\em Annals of Physics}, vol.~118, no.~1, pp.~139--155, 1979.

\bibitem{PhysRevD.86.104017}
P.~Pani, V.~Cardoso, L.~Gualtieri, E.~Berti, and A.~Ishibashi, ``Perturbations of slowly rotating black holes: Massive vector fields in the kerr metric,'' {\em Phys. Rev. D}, vol.~86, p.~104017, Nov 2012.

\bibitem{PhysRevLett.109.131102}
P.~Pani, V.~Cardoso, L.~Gualtieri, E.~Berti, and A.~Ishibashi, ``Black-hole bombs and photon-mass bounds,'' {\em Phys. Rev. Lett.}, vol.~109, p.~131102, Sep 2012.

\bibitem{PhysRevD.96.035019}
M.~Baryakhtar, R.~Lasenby, and M.~Teo, ``Black hole superradiance signatures of ultralight vectors,'' {\em Phys. Rev. D}, vol.~96, p.~035019, Aug 2017.

\bibitem{PhysRevD.88.023514}
R.~Brito, V.~Cardoso, and P.~Pani, ``Massive spin-2 fields on black hole spacetimes: Instability of the schwarzschild and kerr solutions and bounds on the graviton mass,'' {\em Phys. Rev. D}, vol.~88, p.~023514, Jul 2013.

\bibitem{PhysRevD.81.123530}
A.~Arvanitaki, S.~Dimopoulos, S.~Dubovsky, N.~Kaloper, and J.~March-Russell, ``String axiverse,'' {\em Phys. Rev. D}, vol.~81, p.~123530, Jun 2010.

\bibitem{PhysRevD.91.124026}
V.~Cardoso, R.~Brito, and J.~a.~L. Rosa, ``Superradiance in stars,'' {\em Phys. Rev. D}, vol.~91, p.~124026, Jun 2015.

\bibitem{Rao1992AnAP}
S.~Rao, ``An anyon primer,'' {\em arXiv: High Energy Physics - Theory}, 1992.

\bibitem{Sen:1993qc}
D.~Sen, ``{An Introduction to anyons},'' 1993.

\bibitem{Bartolomei:2020qfr}
H.~Bartolomei {\em et~al.}, ``{Fractional statistics in anyon collisions},'' {\em Science}, vol.~368, no.~6487, pp.~173--177, 2020.

\bibitem{KITAEV20032}
A.~Kitaev, ``Fault-tolerant quantum computation by anyons,'' {\em Annals of Physics}, vol.~303, no.~1, pp.~2--30, 2003.

\bibitem{PhysRevLett.69.1849}
M.~Ba\~nados, C.~Teitelboim, and J.~Zanelli, ``Black hole in three-dimensional spacetime,'' {\em Phys. Rev. Lett.}, vol.~69, pp.~1849--1851, Sep 1992.

\bibitem{DAPPIAGGI2018146}
C.~Dappiaggi, H.~R. Ferreira, and C.~A. Herdeiro, ``Superradiance in the btz black hole with robin boundary conditions,'' {\em Physics Letters B}, vol.~778, pp.~146--154, 2018.

\bibitem{PhysRevD.67.084013}
D.~Bini, C.~Cherubini, R.~T. Jantzen, and B.~Mashhoon, ``Massless field perturbations and gravitomagnetism in the kerr-taub-nut spacetime,'' {\em Phys. Rev. D}, vol.~67, p.~084013, Apr 2003.

\bibitem{PhysRevD.61.104013}
C.~Mart\'{\i}nez, C.~Teitelboim, and J.~Zanelli, ``Charged rotating black hole in three spacetime dimensions,'' {\em Phys. Rev. D}, vol.~61, p.~104013, Apr 2000.

\bibitem{Konewko2023ChargeSO}
S.~A. Konewko and E.~Winstanley, ``Charge superradiance on charged btz black holes,'' {\em The European Physical Journal C}, 2023.

\bibitem{PhysRevA.78.063804}
F.~Marino, ``Acoustic black holes in a two-dimensional ``photon fluid'','' {\em Phys. Rev. A}, vol.~78, p.~063804, Dec 2008.

\bibitem{PhysRevD.104.104011}
V.~C, S.~Basak, and S.~Das, ``Hawking radiation of anyons,'' {\em Phys. Rev. D}, vol.~104, p.~104011, Nov 2021.

\bibitem{Cheriyodathillathu:2022fwe}
V.~Cheriyodathillathu, S.~Das, and S.~Basak, ``{Quasinormal modes of anyons},'' {\em Gen. Rel. Grav.}, vol.~56, no.~2, p.~33, 2024.

\bibitem{Visser:1997ux}
M.~Visser, ``{Acoustic black holes: Horizons, ergospheres, and Hawking radiation},'' {\em Class. Quant. Grav.}, vol.~15, pp.~1767--1791, 1998.

\bibitem{SoumenBasak_2003}
S.~Basak and P.~Majumdar, ``‘superresonance’ from a rotating acoustic black hole,'' {\em Classical and Quantum Gravity}, vol.~20, p.~3907, aug 2003.

\bibitem{Brennen2007WhySA}
G.~K. Brennen and J.~K. Pachos, ``Why should anyone care about computing with anyons?,'' {\em Proceedings of the Royal Society A: Mathematical, Physical and Engineering Sciences}, vol.~464, pp.~1 -- 24, 2007.

\end{thebibliography}
\end{document}